\documentclass[showpacs,preprintnumbers,amsmath,amssymb,superscriptaddress,prl,twocolumn]{revtex4}

\def\XXint#1#2#3{{\setbox0=\hbox{$#1{#2#3}{\int}$}
     \vcenter{\hbox{$#2#3$}}\kern-.5\wd0}}

\usepackage{graphicx}
\usepackage{subfigure}
\usepackage{dcolumn}
\usepackage{bm}

\begin{document}

\title{Low-temperature spin Coulomb drag in a two-dimensional electron gas}

\author{A.\ G.\ Yashenkin}

\affiliation{Petersburg Nuclear Physics Institute NRC ``Kurchatov Institute'', Gatchina, 188300
St.~Petersburg, Russia}

\affiliation{\mbox{Department of Physics, Saint Petersburg State University, 7/9 Universitetskaya nab.,
199034 St.~Petersburg, Russia}}

\author{I.\ V.\ Gornyi}

\affiliation{Institut f\"ur Nanotechnologie, Karlsruhe Institute of Technology, 76021 Karlsruhe, Germany}

\affiliation{A.~F. Ioffe Physico-Technical Institute, 194021 St.~Petersburg, Russia}

\affiliation{\mbox{Institut f\"ur Teorie der Kondensierten Materie, Karlsruhe Institute of Technology, 76128 Karlsruhe, Germany}}

\affiliation{L.~D. Landau Institute for Theoretical Physics RAS, 119334 Moscow, Russia}

\date{\today}

\begin{abstract}
The phenomenon of low-temperature spin Coulomb drag in a two-dimensional electron gas is investigated. The spin transresistivity coefficient is essentially enhanced in the diffusive regime, as compared to conventional predictions. The origin of this enhancement is the quantum coherence of spin-up and spin-down
electrons propagating in the same random impurity potential and coupled via the Coulomb interaction. A comprehensive analysis of spin and interlayer Coulomb drag effects is presented.
\end{abstract}

	
\maketitle

{\it Introduction.}
The primary goal of practical spintronics is to manufacture devices capable to manipulate,
storage and transfer the portions of spin polarized quasiparticles in a controlled manner.
This goal can hardly be achieved without detailed understanding of mechanisms of spin decoherence.

The phenomenon of spin Coulomb drag (SCD) \cite{Vig1,Flens-SCD,Vig2,Vig2D,DasSar,Weber} in interacting electron systems is one of these
mechanisms. The Coulomb interaction transfers the momentum between spin-up and spin-down electron
subsystems. This generates the frictional force between these subsystems and the damping of
their relative motion. Due to SCD, the spin packet injected and propagating across a sample in
the external field does not preserve its shape and polarization in time, even in the absence of
spin-flip scattering.

The strength of the SCD effect is characterized by the spin transresistivity $\rho_{\uparrow \downarrow}
 = E_{\uparrow} / J_{\downarrow}$, where $E_{\uparrow}$ is the gradient of the electro-chemical
potential for spin-up electrons and $J_{\downarrow}$ is the current of spin-down electrons, given
the current of spin-up electrons is kept zero. The quantity $E_{\uparrow}$ plays the role of the effective
voltage biased exclusively to spin-up electrons.

So far, the most confident experimental evidence of the SCD effect has been reported by
Weber and co-workers~\cite{Weber}. They observed that the suppression of the spin
diffusion coefficient $D_s$ relative to the charge one $D_c$  could not be interpreted as stemming
exclusively from different Fermi-liquid renormalization of these quantities but
requires also to take into account the SCD,
\begin{equation}
\frac{D_s}{D_c} \propto \frac{1}{1 - \rho_{\uparrow \downarrow}  \, \sigma_0},
\end{equation}
\label{Ds}
where $\sigma_0$ is the total Drude conductivity.

{\it Spin Coulomb drag vs. interlayer Coulomb drag.}
Physically, the phenomenon of SCD is very similar to the one of interlayer Coulomb drag (ICD) in
double layers (see Refs.~ \cite{Pogreb,Jauho,MacDon,Flens-CD,KamOreg,Gramila}; for recent developments, see Ref.~\cite{LevNar} and references
therein). In the conventional ICD setting, the two subsystems participating in the Coulomb-mediated
momentum transfer are quasiparticles located in different layers. Furthermore, the interlayer transresistivity $\rho_{21}$ is
defined as the ratio of the voltage drop in the second (passive) layer to the current passing
through the first (active) layer due to the bias applied in this layer, the current in passive
layer held zero.

Due to this evident similarity, theoretical investigations of ICD and SCD phenomena utilize
common ideas and approaches (cf. Refs.~\cite{Vig1,Flens-SCD,Vig2,Vig2D,DasSar} and Refs.~\cite{Jauho,MacDon,Flens-CD,KamOreg}).
Both interlayer (ITR) and
spin (STR) transresistivity were found to vanish quadratically with lowering temperature, $\rho_{\mu\nu} \propto T^2$;
in certain regimes, the extra factor $\ln T$ appears. The theory developed implies that electrons belonging
to two subsystems propagate in {\it different} random impurity potentials.
As a result, the quadratic temperature dependence of ITR and STR could be interpreted as originating from
coupling of {\it independent} thermal fluctuations of electron density in
different subsystems.

Technically, the principal contribution to $\rho_{\mu\nu}$ in both cases arises from diagrams known
in the condensed matter theory as Aslamazov-Larkin (AL) diagrams (see, e.g., Refs.~\cite{AA,LV}) that, in particular, describe
the fluctuation correction to the conductivity in superconducting systems above the critical temperature.
The main difference between the results for $\rho_{\uparrow \downarrow}$ and $\rho_{21}$ is due to
different couplings between the subsystems which are the intralayer Coulomb interaction between
spin-up and spin-down electrons in the former case and the interlayer one in the latter making
$\rho_{21}$ strongly dependent on the interlayer distance.

Despite of evident similarity of ICD and SCD phenomena, there are essential differences, as well.
The goal of the present paper is to emphasize that they result in a $\it crucially$ distinct low-temperature
(diffusive) behavior of the spin transresistivity, the effect previously overlooked in literature.
More specifically, previous considerations ignored the fact that the spin-up and spin down electrons within
the single layer propagate through the same random impurity potential whereas the impurity potentials
in different layers of a double-layer system are almost uncorrelated in typical experimental setups.

Meanwhile, in Refs.~\cite{GYK,GYK-Ann} a modified experimental setup for the measurement of $\rho_{21}$ was theoretically proposed,
with the impurity potentials for electrons in two layers strongly correlated.
The calculated interlayer transresistivity was found to be strongly enhanced in the
diffusive regime $T \tau < 1$ (here $\tau$ is the electron elastic scattering time; $\hbar = k_B = 1$),
revealing only a weak temperature dependence. On a technical level, the effect of impurity-potential
correlations is described by the class of diagrams associated with the anomalous Maki-Thompson (MT) correction to the conductivity
\cite{LV}.

In the present paper we exploit the ideas of Refs.~\cite{GYK,GYK-Ann} investigating the spin transconductivity
of a two-dimensional electron gas. We find that in the diffusive regime $T \tau < 1$ the spin transresistivity
$\rho_{\uparrow \downarrow}$ is dominated by MT contribution due to the increase (as compared to uncorrelated impurities)
of the interaction time for spin-up and spin-down electrons propagating across the same potential landscape.
We also argue that the MT contribution to STR could be suppressed by applying the external magnetic field.
For clarity, we restrict ourselves to considering the unpolarized
(paramagnetic) case; however, generalization onto polarized electrons is straightforward.

{\it Constituents of the theory}.
We now proceed with details of microscopic calculations.
Let us start from Kubo formula for the
DC spin transconductivity
\begin{equation}
\label{Kubo}
\sigma_{\uparrow \downarrow} = - \left.
\frac{1}{i \omega}\langle\langle \, {\bf J_{\uparrow}}\, ; \, {\bf J_{\downarrow}} \, \rangle \rangle\right|_{\omega \to 0}
\end{equation}
Here the current-current correlation function of the currents of spin-up and spin-down electrons taken in the limit of vanishing external frequency $\omega$
is averaged over realizations of random potential. The spin transresistivity is expressed via the spin conductivity tensor ${\hat \sigma}$ as follows:
\begin{equation}
\rho_{\uparrow \downarrow} = - \frac{\sigma_{\uparrow \downarrow}}{ \sigma_{\uparrow \uparrow} \sigma_{\downarrow \downarrow}- \sigma_{\uparrow \downarrow} \sigma_{\downarrow \uparrow}}.
\end{equation}
As the diagonal components of the matrix ${\hat \sigma}$ contain large and temperature independent Drude
contributions $\sigma_{\mu\mu}^{(0)}=e^2 n_{\mu} \tau / m$, one can approximately set
$\rho_{\uparrow \downarrow} \approx - \sigma_{\uparrow \downarrow}[\sigma_{\uparrow \uparrow}^{(0)} \sigma_{\downarrow \downarrow}^{(0)}]^{-1}.$
Here $\mu= \uparrow , \downarrow$, $n_{\mu}$ is the density of spin-up (spin-down) electrons, and $m$ is the electron mass. We shall consider the paramagnetic  case of equal electron densities $n_{\uparrow} = n_{\downarrow} = n/2$. It yields
$\rho_{\uparrow \downarrow} \approx - 4 \, \sigma_{\uparrow \downarrow} \, \sigma_0^{-2}$.

It is worth commenting on the interaction which couples spin-up and spin-down
electron subsystems. In sharp contrast with the interlayer drag case, the off-diagonal terms
of the Coulomb interaction matrix ${\hat U}$ dressed within the  Random Phase Approximation (RPA)
coincide with the diagonal ones
\begin{equation}
U_{\uparrow \downarrow} ({\bf q}, \omega)= U_{\uparrow \uparrow} ({\bf q}, \omega) = \frac{u_{{\bf q}}}{1- u_{{\bf q}} \left[ \, \Pi_{\uparrow}({\bf q}, \omega) +    \Pi_{\downarrow}({\bf q},\omega) \, \right]}.
\end{equation}
\label{V-RPA}
Here $u_{{\bf q}} = 2 \pi e^2 / {\rm q}$ is the bare Coulomb interaction in two dimension, and
$\Pi_{\uparrow (\downarrow)}({\bf q},\omega)$ is the polarization operator for spin-up (spin-down)
electrons.

{\it Aslamasov-Larkin contribution to spin transresistivity.}
Now we sketch the results for spin transresistivity obtained by neglecting correlations between impurity potentials acting on
spin-up and spin-down electrons. It follows from
Eq.~\eqref{Kubo} that the diagrams for $\sigma_{\uparrow \downarrow}$ consist of two fermionic loops
(one current vertex per each) of electrons with opposite spins.
These loops are connected by off-diagonal elements of the Coulomb interaction matrix ${\hat U} ({\bf q}, \omega)$. To the leading order in
the interaction $U_{\uparrow \downarrow} ({\bf q}, \omega)$, there are two ballistic diagrams
contributing to the spin transconductivity, see Fig.~1a. The low-temperatute renormalization
of these diagrams could be achieved by diffusive decoration of the vertices as it is shown in Fig.~1b.
Notice that no inter-loop impurity lines are drawn as long as the impurity potentials for electrons belonging to different subsystems are considered to be uncorrelated. The resulting (AL) contribution to the transconductivity is given by
\begin{figure}
\includegraphics[scale=0.45]{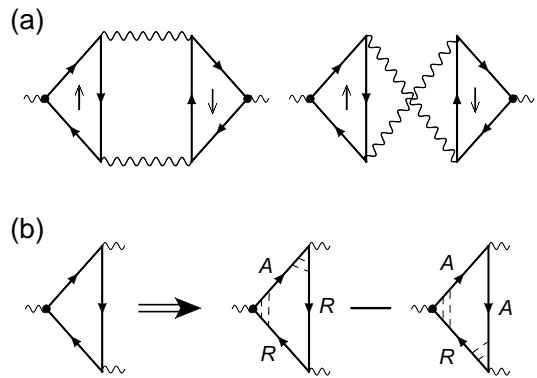}
\caption{
(a) Skeleton diagrams for the Aslamazov-Larkin contribution to the spin transconductivity $\sigma_{\uparrow \downarrow}^{AL}$. Solid lines with arrows depict electron Green's functions and wavy ones stand for the screened interaction $U_{\uparrow \downarrow}$, bold points denote the current (vector) vertices. (b) Diffusive dressing of the AL diagrams, with double-dashed lines depicting the diffuson ladders. Symbols R(A) stand for retarded (advanced) Green's functions. Notice that the dressing of the current vertices is needed only in the case of  {\it smooth} random impurity potentials.
}
\label{fig1}
\end{figure}

\begin{eqnarray}
\label{ALint}
\sigma_{\uparrow \downarrow}^{AL}&=&\frac{2 \pi e^2 \tau^2}{ T \, m^2}\,
\int  (dq)\int (d \omega)\, \frac{ q^2|U_{\uparrow \downarrow} ({\bf q}, \omega) |^2    }{{\rm sinh^2}\frac{\omega}{2T}}
\nonumber \\
  &\times& \,
{\rm Im} \Pi_{\uparrow}({\bf q},\omega) \,  {\rm Im}\, \Pi_{\downarrow}({\bf q},\omega).
\end{eqnarray}
Evaluation of integrals over intermediate frequencies and momenta in Eq.~\eqref{ALint} yields~\cite{Vig2D}
\begin{equation}
\label{ALres}
\rho_{\uparrow \downarrow}^{AL} \approx - \frac{ 2\pi^2 }{3 e^2}\, \left(\frac{T}{\varepsilon_F}\right)^2 \,\left(
\frac{\varkappa_2}{2 k_F} \right)^2
\, F_{\uparrow \downarrow}^{AL} \left(\frac{\varepsilon_F}{T}, \frac{\varkappa_2}{ 2 k_F} \right),
\end{equation}
where $\varepsilon_F$ is the Fermi energy, $k_F$ is the Fermi momentum,  and $\varkappa_2 = 4 \pi e^2 \nu_F$ is the Thomas-Fermi momentum in
two dimensions, with $\nu_F = m / 2 \pi$ being the density of states per spin. The function
$F_{\uparrow \downarrow}^{AL} (x,y) $ defined as
\begin{eqnarray}
\label{AL-F}
F_{\uparrow \downarrow}^{AL} (x,y)
&=&  \frac{\ln x  + \mathcal{C}}{2(1+y)^2}-\frac{1}{(1-y^2)(1+y)}
\nonumber \\   &+& \frac{1+y^2}{(1-y^2)^2} \ln \frac{1+ y}{2 y}
\end{eqnarray}
is a slowly varying function of both its arguments in the entire range of applicability of the theory,
$x < 1$ and $y \lesssim 1$; in particular, it is regular at $y=1$. The numerical constant in Eq.~(\ref{AL-F}) is given by~\cite{Vig2D}: $\mathcal{C}=2+ 2 \ln2-(6/\pi^2)\int_0^\infty dz z^2 \ln z \,\sinh^{-2} z \approx 3.72662.$ For $y\gg 1$ the function (\ref{AL-F}) is expressed in terms of the phenomenological Landau Fermi-liquid interaction parameters.

Equations~\eqref{ALres} and \eqref{AL-F} constitute the main result of the conventional theory \cite{Vig1,Flens-SCD,Vig2,Vig2D,DasSar}.
Notice the quadratic temperature dependence of $\rho_{\uparrow \downarrow}^{AL}$ (with an extra logarithmic
enhancement in certain regimes) as well as the additional smallness of this quantity appearing
in the limit of large electron densities $\varkappa_2 < k_F$.

{\it Maki-Thompson contribution to spin transresistivity.}
Now let us discuss the influence of correlations between the impurity potentials on the motion of interacting quantum particles.
Consider two particles that propagate across the random potential landscape, starting from the
same initial conditions. It is clear that they will keep the relative phase coherence until
one of them experiences the inelastic collision by external scatterer. This is clearly valid also for
particles with opposite spin projection in the absence of spin-orbit scattering.
Therefore, one can expect the quantum interference to maintain
in the channel of (spin) transconductivity, similarly to diagonal components of the conductivity
tensor in a two-component system. The difference is that for the finite transconductivity to exist
the mechanism of momentum transfer between the two subsystems is needed. This is the inelastic
particle-particle interaction. Technically, one should inspect the diagrams with
two fermionic loops interconnected not only by the interaction lines but also by the inter-subsystem
impurity ladders constituting singular diffusons and Cooperons. For such a Cooperon we get
\begin{equation}
\label{Cooperon}
{\rm C}_{\uparrow \downarrow} ({\bf q}, \omega) = \frac{1}{2 \pi \nu_F \tau^2} \, \frac{1}{D q^2 - i \omega + \tau_{\varphi}^{-1} },
\end{equation}
with $\tau_{\varphi}$ being the phase breaking time, $D=\frac{1}{2} {\rm v}_F^2 \tau$ being the
diffusion coefficient, and ${\rm v}_F$ standing for the Fermi velocity. There is no difference
between the ordinary (intra-subsystem) Cooperons and the inter-subsystem ones in the
case of the {\it spin} Coulomb drag because the impurity potential for spin-up
and spin-down electrons is {\it exactly} the same. This is in sharp contrast with the case of
interlayer transport coefficients describing the transport in spatially separated subsystems.
As a result, the interlayer coherence could be achieved only in peculiar experimental geometries, the measure
of the impurity potential correlations being the value of the gap entering interlayer diffusons and Cooperons~\cite{GYK,GYK-Ann}.

\begin{figure}
  \includegraphics[scale=0.45]{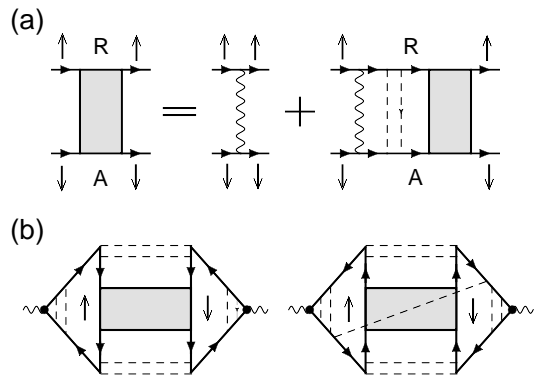}
  \caption{ (a) Diagrammatic (Bethe-Salpeter) equation for the Cooperon-interaction ladder
(shaded box) entering the Maki-Thompson
contribution to the spin transconductivity $\sigma_{\uparrow \downarrow}^{MT}$.
All other notations are the same as in Fig.~1. (b) Maki-Thompson diagrams for spin
transconductivity $\sigma_{\uparrow \downarrow}$ at low temperatures.}
\label{Fig2}
\end{figure}

A closer inspection of additional diagrams reveals strong similarity with the problem of the impurity-correlated
interlayer Coulomb drag. In particular, we observe that the diffuson Hartree diagrams do not contribute to the
spin transconductivity. Furthermore, the series of proper Cooperon diagrams~\cite{footnote} could be summed up
as it is shown in Fig.~\ref{Fig2}; the result is known as the anomalous Maki-Thompson contribution~\cite{AA,LV}.
For the problem considered this contribution to the spin-drag transconductivity has the form
\begin{eqnarray}
\label{MTint}
\sigma_{\uparrow \downarrow}^{MT} &=&  \frac{8 \, e^2 }{T} \,  \int \frac{D \, (dq)}{D q^2 + \tau_{\varphi}^{-1}} \, \int \frac{ (d \omega)}{{\rm sinh^2}\frac{\omega}{2T}} \nonumber \\
  &\times& \,
{\rm Im} \Lambda_{\uparrow\downarrow}({\bf q},\omega) \, {\rm Im}
\Psi_{\uparrow\downarrow}({\bf q},\omega). \end{eqnarray}
The functions $\Psi_{\uparrow\downarrow}$ and $\Lambda_{\uparrow\downarrow}$ in Eq.~\eqref{MTint} are given by
\begin{eqnarray}
\label{lambda-psi}
\Psi_{\uparrow \downarrow} ({\bf q}, \omega) &=& \psi \left[ ( D q^2 - i \omega + \tau_{\varphi}^{-1} + 2 \pi T)/4 \pi T \right]  \nonumber \\
\Lambda_{\uparrow\downarrow}^{-1} ({\bf q}, \omega) &=& \ln \frac{\varepsilon_F}{T}+ \left[ F_{\uparrow\downarrow}^{MT}(\varkappa_2 / 2 k_F ) \right]^{-1} \\
 &-&  \Psi ({\bf q}, \omega) + \psi (1/2),  \nonumber
\end{eqnarray}
where $\psi(z)$ is the digamma function, $F_{\uparrow \downarrow}^{MT} (\varkappa_2 / 2 k_F )=(4 \pi^2 \nu_F)^{-1}
\langle U_{\uparrow \downarrow} ({\bf p} -{\bf p}^{\prime})\rangle_{{\bf p}, {\bf p}^{\prime}}$,
and the angular average over the Fermi surface is given by $\langle ... \rangle_{{\bf p}} = \nu_F
\int_{0}^{2\pi} ... \,\, d \varphi_p$. It is worth noting that due to the large momentum transfer in Eq.~\eqref{MTint}
the interaction $ U_{\uparrow \downarrow} ({\bf q}, \omega)$ could be taken in the static limit,
\begin{equation}
\label{U}
U_{\uparrow \downarrow} ({\bf q}) = \frac{1}{2 \nu_F}\, \frac{\varkappa_2}{\varkappa_2 + {\rm q}}.
\end{equation}
The quantity $F_{\uparrow \downarrow}^{MT} (y)$ is then calculated as follows
\begin{equation}
\label{F^{MT}}
F_{\uparrow \downarrow}^{MT} (y)=\frac{y}{\pi \sqrt{1 - y^2}} \, {\ln} \, \frac{1+\sqrt{1 - y^2}}{1 - \sqrt{1 - y^2}}.
\end{equation}
This RPA expression for $F_{\uparrow \downarrow}^{MT} (y)$ turns out to coincide with the Fermi-liquid interaction
amplitude $F_0^\sigma$ (see, e.g., Ref. \cite{ZNA}). Notice that $F_{\uparrow \downarrow}^{MT} (y\to 0) \propto y \, \ln y$.
In the case of strong interaction $\varkappa_2 / k_F \gg 1$, the function $F_{\uparrow \downarrow}^{MT}$
is given by the Landau Fermi-liquid parameter for the Cooper channel.

Evaluating Eq.~\eqref{MTint} with the use of Eqs.~\eqref{lambda-psi}, \eqref{U} and \eqref{F^{MT}}, we obtain
\begin{equation}
\label{answer}
\rho_{\uparrow \downarrow}^{MT} =  -\frac{2 \pi^2}{3 e^2} \, \frac{\ln T \tau_{\varphi}}{(\varepsilon_F \tau)^2} \,
\left[ \frac{F_{\uparrow \downarrow}^{MT}(\frac{\varkappa_2}{2 k_F})}{1 +\ln \frac{\varepsilon_F}{T} \, F_{\uparrow \downarrow}^{MT}(\frac{\varkappa_2}{2 k_F})}   \right]^2.
\end{equation}
In the diffusive limit, the phase-breaking time is dominated by the Coulomb
interaction and is given in two dimensions by
$\tau_\varphi \sim g \,(T \ln g)^{-1}$,
where $g=2\varepsilon_F\tau$ is the dimensionless conductance~\cite{AA}.
The anomalous Maki-Thompson contribution to the transresistivity, Eq.~\eqref{answer}, constitutes our main
result. We note that this correction is proportional to the \textit{squared} effective Cooper-channel interaction amplitude \cite{Fin,BGM}
at scale $T$. This can be traced back to the fact that the contribution of diagrams
with a {\it single}  interaction line connecting two fermionic loops vanishes identically for an arbitrary interaction already
before disorder averaging \cite{ZNA}.

\begin{figure}
\includegraphics[scale=0.7]{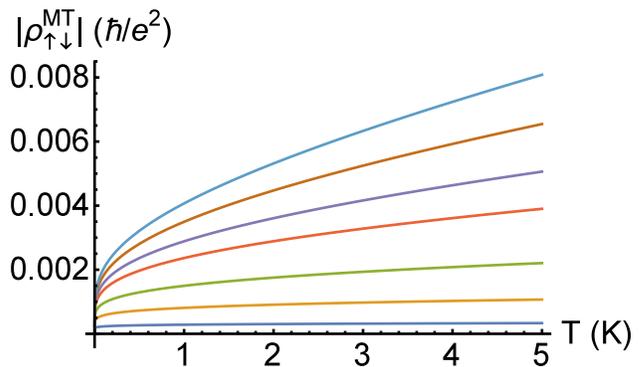}
\caption{
The absolute value of the Maki-Thompson contribution to the spin-drag transresistivity, Eq.~(\ref{answer}), shown as a function of $T$ up to
$T=1/\tau=5$K for $\varepsilon_F=50$K, $g=20$, and $\varkappa_{2} /k_F=0.02,\ 0.1,\ 0.2,\ 0.4,\ 0.6,\ 1,\ 1.8$ (from bottom to top).
}
\label{fig:rhoMT}
\end{figure}

{\it Discussion and conclusions.}
Remarkably, since the factor $\ln \,(T\tau_\varphi)\sim \ln g$ is independent of $T$, the temperature dependence
of $\rho_{\uparrow \downarrow}^{MT}$ is only due to $\ln \, (\varepsilon_F/T)$ in the denominator of Eq.~(\ref{answer}),
where it is multiplied by the interaction constant.
Thus the STR reveals the phenomenon of (quasi)saturation: for not too strong interactions it stays almost constant
down to $T\sim \varepsilon_F \exp \, (-1/F_{\uparrow \downarrow}^{MT})$;
still in the limit $T\to 0$ the SCD vanishes as $\ln^{-2} T $.
The anomalous Maki-Thompson contribution to $\rho_{\uparrow \downarrow}$ for $T\tau<1$ is presented in Fig.~\ref{fig:rhoMT}.

\begin{figure}
\includegraphics[scale=0.7]{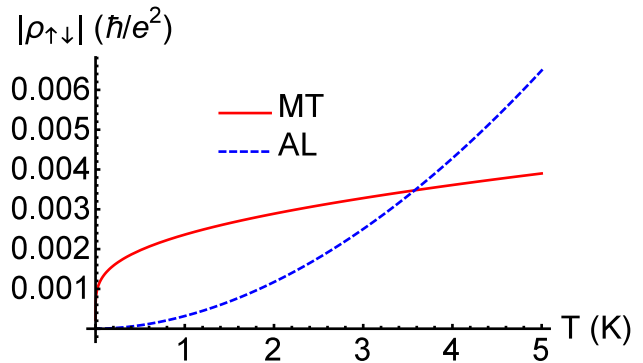}
\caption{
The comparison of the Maki-Thompson (red solid curve) and Aslamazov-Larkin (blue dashed curve) contributions to the spin-drag transresistivity
at low temperatures $T<1/\tau=5$K
for $\varepsilon_F=50$K, $g=20$, and $\varkappa_{2} /k_F=0.4$.
}
\label{fig:compare}
\end{figure}

 Let us compare our result given by Eq.~\eqref{answer} with the result of conventional theory.
For simplicity, we omit all weak logarithmic dependencies. Comparing then the AL and MT contributions, we observe the qualitative similarity
in their dependencies on parameter $\varkappa_2 / k_F$. The rest of the formulas provides us with the estimate
\begin{equation}
\label{interval}
T < \tau^{-1}
\end{equation}
for
temperature interval wherein the {\it anomalous Maki-Thompson contribution to the STR dominates over the Aslamazov-Larkin one},
$\left|\rho_{\uparrow\downarrow}^{MT}\right| > \left|\rho_{\uparrow\downarrow}^{Al}\right|$, see Fig.~\ref{fig:compare}. This region coincides with
the range of applicability of Eq.~\eqref{answer} evaluated within the diffusion approximation.
Specifically, while the conventional AL-type contribution vanishes quadratically with decreasing temperature, the MT contribution
stays almost constant down to exponentially low temperatures.

In this paper, we have restricted our analysis with the paramagnetic situation. The generalization onto
polarized electrons in straightforward. In particular, the quantity $F_{\uparrow \downarrow}^{MT}$
acquires dependence on electron magnetization and the Cooperon, Eq.~(\ref{Cooperon}), acquires
a frequency shift $\omega_M = (n_\uparrow-n_\downarrow) \, \nu_{F}^{-1}$.
This leads to a suppression of the Maki-Thompson contribution; however, it still dominates the spin
drag,  provided $\omega_M \tau < 1$. Furthermore, it is straightforward to extend our consideration
onto other dimensions, as well as to the case of finite magnetic field. In the latter case, we
announce the suppression of the MT contribution to STR due to suppression of Cooperons. These
results will be published elsewhere~\cite{YG}.

In conclusion, we have investigated the spin Coulomb drag effect at low temperatures.
We observed the novel contribution to the spin transresistivity coefficient which is due to maintaining of
quantum coherence between the electrons with spin-up and spin-down propagating in the {\it same random impurity potential}
and interacting with each others via the Coulomb interaction. The diagrams responsible for this process are similar to those
describing the anomalous Maki-Thompson corrections to the conductivity. We argue that this mechanism dominates
in the entire diffusive temperature interval
leading to a dramatic enhancement of the spin Coulomb drag as compared to previous studies.

{\it Acknowledgments.} Authors benefited from previous discussions with D.V.\ Khveshchenko.
This work is supported by Russian Science Foundation Grant No.\ 14-22-00281 (AGY).


\begin{thebibliography}{100}
\bibitem{footnote} We note that the diagrams describing the density-of-states correction
in the Cooper channel~\cite{LV} do not contribute to the transresistivity because the current vertices in these diagrams
are inserted into the same fermionic loop, thus giving rise to corrections to $\sigma_{\mu\mu}$ only.
Furthermore, all diagrams with current vertices inserted into two distinct fermionic loops without changing
retarded Green's functions to advanced at current vertices vanish due to
gauge invariance \cite{AALK,GYK}.
\end{thebibliography}
\end{document}